\documentstyle[12pt,aasms4]{article}

\def\deg{\ifmmode^\circ\else$^\circ$\fi}

\def\kpc{\ifmmode h^{-1}{\rm kpc}\else$h^{-1}{\rm kpc}$\fi}
\def\kms{\ifmmode {\rm km~s}^{-1}\else${\rm km~s}^{-1}$\fi}
\def\bii{\ifmmode b^II\else b$^{II}$\fi}
\def\lii{\ifmmode l^II\else l$^{II}$\fi}
\def\feh{\ifmmode {\rm [Fe/H]}\else [Fe/H]\fi}

\lefthead{Beers et al.}
\righthead{Metal-Poor Giants from the LSE Survey}

\begin{document}

\title{Metal Abundances and Kinematics of Bright Metal-Poor Giants Selected
from the LSE Survey:  Implications for the Metal-Weak Thick Disk}

\author{Timothy C. Beers\altaffilmark{1,2}}
\affil{Department of Physics \& Astronomy, Michigan State University, E.
Lansing, MI 48824\\email:  beers@pa.msu.edu}
\author{John S. Drilling\altaffilmark{1}}
\affil{Department of Physics \& Astronomy, Louisiana State University, Baton
Rouge, LA  70803\\email:  drilling@rouge.phys.lse.edu}
\author{Silvia Rossi\altaffilmark{1,2}}
\affil{Instituto Astron\^omico e Geof\'isico, Universidade de S\~ao Paulo,
Brazil\\email: rossi@orion.iagusp.usp.br}
\author{Masashi Chiba}
\affil{National Astronomical Observatory, Mitaka, Tokyo 181-8588, Japan\\
email: chibams@gala.mtk.nao.ac.jp}
\author{Jaehyon Rhee}
\affil{Department of Astronomy, University of Virginia, Charlottesville, VA 22903\\
email:  rhee@virginia.edu}
\author{Birgit F\"uhrmeister\altaffilmark{2}}
\affil{Hamburger Sternwarte, Universit\"at Hamburg, Gojensbergsweg 112,
D-21029, Hamburg, Germany\\email:  bfuhrmeister@hs.uni-hamburg.de}
\author{John E. Norris}
\affil{Research School of Astronomy \& Astrophysics, Australian National
University, Mount Stromlo Observatory, Cotter Road, Weston ACT 2611, Australia;
jen@mso.anu.edu.au}
\and
\author{Ted von Hippel}
\affil{Department of Astronomy, University of Texas, Austin, TX  78712;
ted@astro.as.utexas.edu} 

\altaffiltext{1}{Visiting Astronomer, Cerro Tololo
Inter-American Observatory, which is operated by the Associated Universities
for Research in Astronomy, Inc., under contract with the National Science
Foundation.}
\altaffiltext{2}{Visiting Astronomer, the European Southern Observatory.}

\begin{abstract}

We report medium-resolution (1--2 \AA ) spectroscopy and broadband ($UBV$)
photometry for a sample of 39 bright stars (the majority of which are likely to
be giants) selected as metal-deficient candidates from an objective-prism
survey concentrating on Galactic latitudes below $|b| = 30\deg$, the LSE survey
of Drilling \& Bergeron.  Although the primary purpose of the LSE survey was to
select OB stars (hence the concentration on low latitudes), the small number of
bright metal-deficient giant candidates noted during this survey provide
interesting information on the metal-weak thick disk (MWTD) population.  Metal
abundance estimates are obtained from several different techniques and
calibrations, including some that make use of the available photometry and
spectroscopy, and others that use only the spectroscopy; these methods produce
abundance estimates that are consistent with one another, and should be secure.
All of the targets in our study have available high-quality proper motions from
the {\it Hipparcos} or {\it Tycho-II} catalogs, or both, that we combine with
radial velocities from our spectroscopy to obtain full space motions for the
entire sample.

The rotational (V$_{\phi}$) velocities of the LSE giants indicate the presence
of a rapidly rotating population, even at quite low metallicity.  We consider
the distribution of orbital eccentricity of the LSE giants as a function of
[Fe/H], and conclude that the local fraction (i.e., within 1 kpc from the Sun)
of metal-poor stars that might be associated with the MWTD is on the order of
30\%--40\% at abundances below [Fe/H] $= -1.0$.  Contrary to recent analyses of
previous (much larger) samples of non-kinematically selected metal-poor stars
(assembled primarily from prism surveys that concentrated on latitudes above
$|b| = 30 \deg$), we find that this relatively high fraction of local
metal-poor stars associated with the MWTD may extend to metallicities {\it
below} [Fe/H] = $-1.6$, much lower than had been considered before.  We
identify a subsample of 11 LSE stars that are very likely to be members of the
MWTD, based on their derived kinematics; the lowest metallicity among these
stars is [Fe/H] $= -2.35$.  Implications of these results for the origin of the
MWTD and for the formation of the Galaxy are considered.

\end{abstract}


\section{Introduction}

Although considerable efforts have been made over the past few decades to
identify metal-deficient stars in the Galaxy, there remains a dearth of
recognized metal-poor giants in the solar neighborhood, particularly those
located close to the Galactic plane.  Indeed, until quite recently it was
assumed that the metallicity distribution function of the thick-disk component
of the Galaxy cut off rather sharply below [Fe/H] $ \approx -1$, hence the only
expected contributor to a local metal-weak population of giants would be the
extremely low density halo population.  Even if one takes the view that such
metal-weak stars might exist in the solar neighborhood, there are clear reasons
why they might have been heretofore overlooked: (1) The selection criteria for
most surveys of (non-kinematically selected) metal-poor stars begins by
concentrating on areas of the sky above Galactic latitude $|b| = 30\deg$, so as
to minimize the number of spurious candidates included from the more metal-rich
(and much higher density) disk populations (thick and thin), (2) Recent
objective-prism surveys have concentrated on fainter targets, and generally
saturate at brighter apparent magnitudes, and (3) Though one might have hoped
to find nearby (bright) metal-poor stars amongst high proper-motion catalogs,
if a significant fraction of local metal-weak stars possess kinematics of a
disk-like population, they will have been selected against in these catalogs.
Even when one considers high Galactic latitudes, there does not exist a
plethora of recognized nearby metal-poor giants.  For example, there are only
32 bright giants with $\feh \le -2.0$ in the recent study of Burris et al.
(2000), essentially all drawn from the objective-prism survey of Bond (1980).
The Beers et al.  (2000) catalog (based on a compilation of numerous sources)
lists only 75 giants with $\feh \le -2.0$ and with $V \le 12.0$.

The detection of relatively nearby metal-poor stars would comprise a useful
sample for many investigations.  For example, metal-poor stars near the disk
plane are a-priori much more likely to be members of the metal-weak thick-disk
(hereafter, MWTD) population\footnote{It remains unclear whether the MWTD (with
a low-metallicity tail extending down to at least [Fe/H] $= -1.6$, and as we
argue in this paper, probably lower) is properly considered a separate
population from the canonical thick disk (with a metallicity distribution
function peaking around [Fe/H] $\sim -0.6$), or if it is in reality the
metal-weak tail of this same population; for simplicity of the nomenclature, we
refer to the MWTD as an individual population, though we hope to address its
relationship to the canonical thick disk based on new and more extensive
surveys in the near future.}, which several authors have argued includes stars
as metal-deficient as $\feh \sim -1.6$ (Norris, Bessell, \& Pickles 1985;
Morrison, Flynn, \& Freeman 1990, hereafter MFF; Morrison 1993; Beers \&
Sommer-Larson 1995; Layden 1995; Martin \& Morrison 1998; Chiba, Yoshii, \&
Beers 1999; Katz et al. 1999; Chiba \& Beers 2000), and perhaps even lower.
One of the motivations for the present work was to test whether the relative
fraction of MWTD stars in a sample of bright metal-poor giants located near the
Galactic plane might be substantially higher than previously claimed, owing to
the low-latitude cutoffs of most kinematically unbiased surveys.

The general pattern of relative elemental abundances for stars thought to be
members of the MWTD population is still poorly known, although recent efforts
are improving the situation (Fuhrmann 1998; Bonifacio, Centurion, \& Molaro
1999; Mashonkina \& Gehren 2000; Prochaska et al. 2000).  Because of their
lower temperatures, metal-deficient giants have much richer absorption-line
spectra than their warmer main-sequence counterparts, providing the opportunity
to study many more elemental species (e.g., Burris et al. 2000; Norris, Ryan,
\& Beers 2001).  In addition, with the completion of the {\it Hipparcos}
mission (ESA 1997), and the recently released {\it Tycho-II} catalog (Hog et
al.  2000), many stars brighter than $V \sim 12$ now have accurately measured
proper motions, allowing for the derivation of full space motions, once radial
velocities are obtained and distance estimates are made.  Clearly, efforts to
increase the number of recognized bright metal-poor giants are important.

The original Case-Hamburg OB-Star surveys (see Stephenson \& Sanduleak 1971,
and references therein) primarily concentrated on Galactic latitudes within the
relatively narrow region $-10\deg \le b \le +10\deg$.  The Luminous Stars
Extension (LSE) survey of Drilling \& Bergeron (1995) sought to detect
additional OB stars (in particular extreme helium stars and very hot OB
subdwarfs) by extending the original Case-Hamburg surveys to cover the Galactic
latitude range $b = \pm 10\deg $ to $b = \pm 30\deg$ in the Galactic longitude
interval $-60\deg \le l \le +60\deg$\footnote{ A portion of this range was
intersected by plates taken in connection with the LS IV survey; these regions
were {\it not} inspected.  See Fig. 1 of Drilling \& Bergeron (1995).}.  In the
course of this effort, a number of apparently metal-deficient late-type stars,
most of which were expected to be giants, were noted in the process of visual
inspection of the objective-prism plates.

In this paper we report new medium-resolution (1--2 \AA ) spectroscopy for all
39 candidate metal-poor giants from the LSE survey, and for the majority of the
sample, newly measured broadband $UBV$ photometry.  In \S2 we describe the
acquisition of the spectroscopy, the measurement of radial velocities and
line-strength indices, the newly obtained broadband photometry, and reddening
and distance estimates.  Estimation of reddening is more important for the
present sample of stars than for stars with $|b| > 30\deg$, owing to the
generally higher values of color excess, and the increase in the patchiness of
interstellar dust and gas at lower latitudes.  As such, we seek to find
consistency between estimates of de-reddened colors that make use of measured
photometry and independent estimates of de-reddened color from a newly defined
Balmer-line index.  We then apply several separate approaches to obtain
estimates of the metallicities of our program stars, including the calibration
of Beers et al. (1999), a newly calibrated artificial neural network
(hereafter, ANN) approach based on line-index information, as well as a
previously calibrated ANN approach (Snider et al. 2001) that makes use of the
full set of input pixels of each program spectrum.  In \S 3 we report {\it
Hipparcos} and {\it Tycho-II} proper motions, and describe the derivation of
space motions for the LSE stars.  We then consider the kinematics of the LSE
giants, in particular their rotational velocities, and compare them with those
of other bright metal-poor giants with space motions provided by Chiba \& Beers
(2000).  The distribution of derived orbital eccentricities is then used to
consider the fraction of MWTD stars that are represented in this new sample.  A
summary of our results, and a discussion of their implications, are presented
in \S4.

\section{Spectroscopy, Radial Velocities, Photometry, and Distance Estimates}

\subsection{Spectroscopic Measurements and Data Reduction}

The LSE metal-deficient candidates observed in our program (designated as
``MD?'' in the original spectroscopic classifications of Drilling \& Bergeron
1995) are provided in Table 1.  Column (1) lists the star name.  Columns (2)
and (3) list the (J2000.0) equatorial coordinates of the stars.  The Galactic
longitude and latitude for each star is listed in columns (4) and (5),
respectively.  The approximate $V$- or $B$-band apparent magnitude, as provided
in the HST Guide Star Catalog, is listed in column (6), with the appropriate
band noted.

Most of the LSE candidates were observed as ``fillers'' during other
spectroscopic campaigns, when conditions were less than optimal for the primary
program.  As a result, the medium-resolution (1-2 \AA\ over 2 pixels)
spectroscopy reported in this paper has been obtained using a number of
telescopes and instrumentation.  Table 2 lists the telescopes, detectors,
wavelength coverage, dispersion of the spectra, and the numbers of stars
observed with each combination of equipment.  The source of the spectroscopic
data for each star is indicated by the code in column (7) of Table 1.

The LSE stars were typically observed to a minimum signal-to-noise (S/N) ratio
of approximately 20/1 at 4000 \AA .  In a number of cases much higher S/N
spectra were obtained.  Spectra of calibration arc lamps were obtained before
or after each program star, and nightly flatfields and bias frames were taken.
Data reduction followed standard procedures using the IRAF\footnote{IRAF is
distributed by the National Optical Astronomy Observatories, which are operated
by the Association of Universities for Research in Astronomy, Inc., under
cooperative agreement with the National Science Foundation.} suite of routines
as described in Beers et al. (1999).  Figure 1 shows several example spectra of
metal-deficient LSE candidates with similar colors, arranged from relatively
metal-rich to relatively metal-poor.

\subsection{Measurement of Radial Velocities and Line Indices}

Radial velocities were measured for each of our program stars using the
line-by-line and cross-correlation techniques, described in detail by Beers et
al. (1999) and references therein.  The spectral resolution is similar to that
obtained for the majority of the HK survey follow-up, hence we anticipate that
the measured radial velocities should be accurate to the same level, on the
order of 7--10 \kms (one-sigma) .  Comparison with radial velocities for
standard stars observed during the same campaigns during which our program was
conducted (and with similar signal-to-noise ratios as our program objects)
indicate that this accuracy was indeed achieved.  A few of these stars have had
high-resolution measurements obtained during the course of the Cayrel et al.
Large Programme with VLT/UVES -- all velocities are consistent within the above
quoted one-sigma error.  Measurements of heliocentric radial velocities, after
correction for the Earth's rotation and orbital motion, are listed in column
(8) of Table 1.  Published radial velocities, based on high-resolution
spectroscopy for two of our stars, provide additional confidence that our
velocity measurements are within the expected errors.  For LSE-149 (HD~178443),
Bond (1980) obtained V$_{\rm rad} $ = 102 \kms , which differs by only 4 \kms\
from the value reported in Table 8. For LSE-182 (HD~184711), McWilliam et al.
(1995a) report V$_{\rm rad}$ = 343 \kms, identical to the value reported in
Table 8.

For each star, the derived (geocentric) radial velocities were used to place a
set of fixed bands for the derivation of line-strength indices, which are
pseudo-equivalent widths of prominent spectral features.  The bands we employ
are summarized in Table 3.  A complete discussion of the choice of bands, and
the ``band-switching'' scheme used to produce our derived Ca {\sc II} K-line
index, $KP$, and the Balmer-line index, $HP2$, which measures the strength of
the H-$\delta$ line, is provided in Beers et al. (1999).  The additional
Balmer-line index, $HG2$, is a band-switched measurement of the strength of the
H-$\gamma$ line, and is defined in a completely analogous manner to $HP2$.

Line indices (in \AA\ ) for prominent spectral features for each of the
program stars are reported in columns (9) -- (14) of Table 1.  Based on
repeated measurements of numerous standard stars, our expectation is that, for
a spectrum of reasonably good S/N ratio (S/N $= 20$ or more) ,
errors in the line indices on the order of 0.1 \AA\ are achieved.  In order for
a line-index measurement to be considered a detection, we require that the
derived indices be above a minimum value of 0.25 \AA.  Indices that failed to
reach this minimum value are indicated in the table as missing data.

In addition to the line-strength indices, we have measured an Auto-Correlation
Function index for each spectrum, as described in detail in Beers et al.
(1999).  We actually make use of the base-10 logarithm of this index, hence
it is referred to as $LACF$.  The $LACF$ index quantifies the strength of the
multitude of weak metallic lines that are present in each spectrum, and
provides an additional indicator of the overall abundance.  It is of particular
use for cooler stars, such as many of those in the present program, where the
primary metallicity indicator we employ (the CaII K-line $KP$ index) approaches
saturation for stars with [Fe/H] $ > -1.0$.  The LSE spectra were obtained
with a variety of resolutions, hence appropriate correction factors were
applied to bring them onto a common system.  The calibration procedure of Beers
et al.  (1999) obtains an optimum metallicity estimate by consideration of both
the $KP$ index and the $LACF$ index at a given color.  As described below, we
also make use of the $LACF$ in the training of ANNs to derive metallicity
estimates.

\subsection{Broad-band $UBV$ Photometry and Reddening Estimation}

\subsubsection{Newly Obtained $UBV$}

Previously unpublished $UBV$ photometry for 20 of our 39 targets was obtained
with the 0.9 m telescope at Cerro Tololo Inter-American Observatory, on the
nights of 1980 July 10, 11, and 13, using a standard photoelectric photometer
and filters. The reduction procedure outlined by Schulte \& Crawford (1961) was
used, adopting the following mean extinction coefficients: k = 0.15, k$_1$ =
0.10, k$_2$ = $-$0.03, k$_3$ = 0.32 and k$_4$ = 0.00. Dead times,
transformation coefficients, and night corrections were determined from 55
observations of standard stars for which magnitudes and colors are given by
Johnson (1963), Johnson et al. (1966), and Landolt (1973).  These stars were
observed over the same range in color, airmass, and declination as our program
stars. Any systematic differences are small compared to the random mean errors:
$\sigma V = 0.014$ mags, $\sigma (B-V) =  0.011$ mags, and $\sigma
(U-B) = 0.016$ mags, respectively, for a single observation.

Table 4 lists the new photometry, as well as photometry reported
in the SIMBAD database and taken from the {\it Tycho-II} catalog.  There are
twelve stars in Table 4 for which photometry was obtained from the SIMBAD
database, and seven stars for which photometry was taken from the {\it
Tycho-II} catalog.  Note that the errors in the {\it Tycho-II} photometry can
become quite large ($> 0.15$ mags) for the stars with $V > 10.5$ (Hog et al.
2000), so improved photometry should be obtained for these stars in the near
future.  Note, however, that for stars with colors $(B-V)_0 \ge 0.7$, the
dependence of two of the metallicity indicators we employ (the $KP$ and $LACF$
indices) on the measured color is not very strong, so modest errors in the
derived colors can be tolerated.  Nevertheless, as described below, we carry
out several checks on the appropriate colors to apply in subsequent analysis of
this data.  Also note that, as addressed below, the trained ANNs make use
exclusively of spectral information, and hence are not subject to metallicity
errors arising from poor photometry.

\subsubsection {Reddening and Distance Estimates}

Because the LSE metal-poor candidates all have $|b| < 30\deg$, careful
attention must be paid to the reddening corrections.  We initially adopted the
Schlegel, Finkbeiner, \& Davis (1998) estimates of reddening listed in column
(2) of Table 5. The Schlegel et al. estimates have superior spatial resolution,
and are thought to have a better-determined zero point, than the Burstein \&
Heiles (1982) maps.  However, Arce \& Goodman (1999) caution that the Schlegel
et al.  map may overestimate the reddening values when their reported color
excess, $E(B-V)_S$, exceeds about 0.15 mags.  Our own independent tests
suggest that this problem may extend to even lower color excesses, on the order
of $E(B-V)_S = 0.10$ mags.  Hence, we have adopted a slight revision of the
Schlegel et al. reddening estimates, according to the following:

\begin{equation}
\begin{array}{lclcl}
E(B-V)_A & = & E(B-V)_S &\;\;\;\;\; & E(B-V)_S \le 0.10\\ [.25in]
E(B-V)_A & = & 0.10 + 0.65 \times [E(B-V)_S - 0.10] &\;\;\;\;\;&  E\;(B-V)_S > 0.10
\end{array}
\end{equation}

\noindent where $E(B-V)_A$ indicates the adopted reddening estimate.  We note
that for $E(B-V)_S \ge 0.15$ this approximately reproduces the 30\%--50\%
reddening reduction recommended by Arce \& Goodman (1999).  To account for
stars that are located within the reddening layer, assumed to have a scale
height $h = 125$ pc,  the reddening to a given star at distance $D$ is reduced
compared to the total reddening by a factor $[1-\exp(-|D\; \sin\; b|/h)]$.

Distances to individual stars are estimated from $M_V ~~vs. ~~ (B-V)_0$
relations, as described in Beers et al. (2000).  The procedure must be
iterated, because both $V_0$ (and therefore $D$) and  $(B-V)_0$ depend on the
adopted reddening.  Since the $M_V ~~vs. ~~ (B-V)_0$ relations depend on
metallicity, as well as on the classification of the star, at each step of the
iteration the metallicity is re-computed and the classifications re-determined
with the current estimates of $(B-V)_0$ and $(U-B)_0$, so that at the end we
obtain consistent estimates of the final reddening, $E(B-V)_F$, $D$, and
[Fe/H].  Based on the work of Beers et al. (2000), we estimate that these
distances should be accurate to approximately 10-20\%, although in cases of
highly reddened individual stars, they may exceed 20\%.  We consider
the impact of distance errors on the derived kinematics of our program stars
in \S 3.2 below.

Fortunately, we are not required to rely solely on photometric estimates of the
intrinsic colors and reddening, as the line strengths of the observed Balmer
lines also provide a means by which a de-reddened color may be derived.  To
implement these estimates, we have trained an ANN (using the commercially
available ``Backpack 4.1'' routine, from Zsolutions.com), taking as inputs the
base-10 logarithm of the mean Balmer-line index, $\log [(HP2 + HG2)/2]$ (which
we refer to as $LDGP$ below), the logarithm of the $KP$ index ($LKP$), and the
logarithm of the ACF ($LACF$), and producing as output an estimate of the
intrinsic color, which we refer to as $BV_{ANN}$.  For general comments about
the use of ANNs for problems of this sort, see the extensive discussion in
Snider et al. (2001).

The training of the color-estimation ANN was carried out using the subset of
398 of the 551 ``standard stars'' described by Beers et al.  (1999) for which
measures of all three inputs were available, setting aside 20\% of this sample
for use as a validation set to estimate errors in the procedure.  Experiments
with the number of hidden nodes indicated that minimum errors were obtained
with the use of no more than six hidden nodes arranged in a single
layer. \footnote {In an ANN with a single hidden layer, such as presented here,
each node in the hidden layer receives the normalized sum of the weighted
inputs, ${1 \over N} \Sigma w_{ij} (input)$.  Each hidden node performs a
non-linear operation on its input, allowing the input data to be transformed
to a set of non-linear parameters, the number of which is equal to the number
of hidden nodes.  These parameters, the outputs of the hidden nodes, are then
multiplied by the weights, summed, and normalized, at which point the result of
the ANNs is the desired physical parameter, or classification, of a given star.
The training procedure is an iterative process of automatically adjusting the
weights to minimize the classification error.}  The overall one-sigma error in
prediction of $(B-V)_0$ obtained over the color range $0.3 \le (B-V)_0 \le 1.2$
was $\sigma (B-V)_0 = 0.054$ mags, with a median offset in estimated color of
+0.004 mags.  Note, however, that the size of the estimated errors is rather
different in the color ranges $0.3 \le (B-V)_0 \le 0.8$ and $0.8 < (B-V)_0 \le
1.2$.  For the bluer stars, a prediction error of $\sigma (B-V)_0 = 0.047$ mags
was achieved, while for the redder stars, the errors degraded to $\sigma
(B-V)_0 = 0.122$ mags.  For both ranges the median color offsets remained
small, on the order of 0.003 mags.  Estimates of de-reddened $(B-V)_0$ colors
obtained by the ANN approach, $BV_{ANN}$, are listed in column (8) of Table 5.

For convenience, in Table 5 we have also listed the measured $B-V$ colors
and their sources, in columns (2) and (3), respectively.  Column (4) lists the
initial reddening from Schlegel et al. (1998), $E(B-V)_S$, while column
(5) lists the adopted initial reddening, after reduction in some cases as
described above, $E(B-V)_A$.  The first-pass distance-corrected
estimate of reddening obtained from the iterative procedure described above,
$E(B-V)_F$, is listed in column (6); the resulting first-pass de-reddened
color $(B-V)_0$ is listed in column (7).  Comparison of the first-pass
de-reddened colors in column (7) with the ANN estimates listed in column (8)
reveals general agreement, at least for stars with measured de-reddened colors
in the range $(B-V)_0 \le 1.0$.  For the 17 stars in this color range with
photometry in which we have the greatest confidence (listed as source ``P''),
the median offset in $BV_{ANN} - (B-V)_0$ is $-$0.050 mags, with a one-sigma
scatter between the two estimates of de-reddened color of $\sigma = 0.067$
mags.  For the 17 stars where photometry is drawn from either the SIMBAD
database or the {\it Tycho-II} catalog, which are likely to have larger errors,
the median offset between the de-reddened color estimates in this same range of
color is $-0.010$ mags, with a one-sigma scatter of $\sigma = 0.074$ mags.

There is no guarantee that the final Schlegel et al. estimates of reddening
listed in column (6) of Table 5 are themselves correct, so we have decided to
proceed, for stars with $(B-V)_0 \le 1.0$, using a straight mean of the two
estimates of de-reddened color listed in columns (7) and (8).  The mean value
of estimated de-reddened color is listed in column (9), and is designated
$<(B-V)_{0}>$.  Since, for stars with $(B-V)_0 > 1.0$, the $LDGP$ index is
quite small, and subject to greater observational errors reflecting the
weakness of the Balmer lines upon which it is based, we are concerned about the
accuracy of the listed $BV_{ANN}$ estimates for a few of the program stars.  In
these cases we have simply adopted the value obtained from the photometric
estimate listed in column (7).  One can then define an ``effective reddening,''
$E(B-V)_E = (B-V) - <(B-V)_{0}>$, which we list in column (10) of Table 5.  In
some cases, this effective reddening is less than zero, due to possible errors
in the reported colors of stars for which we have not obtained measured
photometry of our own.

We proceed with the type classifications, estimated absolute magnitudes, and
associated distance estimates, carried out according to the procedures
described by Beers et al. (2000), based on our best estimates of de-reddened
colors, $<(B-V)_{0>}$, and reddening, $E(B-V)_E$, as obtained above.  The
assigned classification of each star is listed in column (11) of Table 5.
Columns (12) and (13) list the adopted absolute magnitude and distance
estimates, respectively.

\subsection{Metallicity Estimates}

Much of the past debate concerning the reality of the MWTD has centered around
the validity of estimated stellar abundances for putative members of this
population (e.g., Twarog \& Anthony-Twarog 1994; Ryan \& Lambert 1995).  Hence,
we have endeavored to take particular care in the present study to obtain
metallicity estimates from several different approaches.  Broadly speaking, we
can divide the methods we employ into two categories, ``photometric'' abundance
estimates, which involve the use of line indices and estimates of de-reddened
$(B-V)_0$ colors, and ``non-photometric'' abundance estimates, which make use
of line indices or spectral information that {\it does not} depend on estimates
of de-reddened colors, and thus provides some confidence that a grossly
incorrect metallicity is not derived as the result of an incorrectly adopted
de-reddened color.  We have also used a number of different calibrations (all
of which are based on subsets of the Beers et al.  1999 standard stars) to
ensure that our final results are not dependent on any single calibration.  The
two sets of estimation procedures are discussed below.

\subsubsection{Estimates Using Estimated De-reddened Colors}

Beers et al. (1999) describe a technique for the estimation of [Fe/H] from
medium-resolution spectroscopy of stars, based on the strength of the Ca {\sc
II} K-line index, $KP$, and the $LACF$ index, as a function of de-reddened
$(B-V)_0$ color, with accuracy on the order of 0.15--0.2 dex over the abundance
range $-4.0 \le \feh \le +0.3$. This method makes use of an optimal combination
of independent estimates obtained from the $KP$ line indices and those obtained
from the $LACF$ measurements, based on comparisons with predictions of these
quantities from synthetic spectra and colors, constrained by observations of a
large set of standards with available external high-quality abundance
estimates.  In Table 6 we list the results of these calculations.  Column (1)
lists the star name, while column (2) lists the estimated metallicity obtained
by application of the Beers et al. (1999) procedure, [Fe/H]$_{\rm AK2}$, and
its associated one-sigma error.

As an alternative, we have trained an ANN, taking as inputs $LKP$, $LACF$, and
the de-reddened color estimate $(B-V)_{0}$, and producing as output an
estimate of the metallicity [Fe/H].  The training was carried out using the
subset of 405 of the 551 ``standard stars'' described by Beers et al. (1999)
for which measures of all three inputs were available, setting aside 20\% of
this sample for use as a validation set to estimate errors in the procedure.
As we found in the ANN prediction of de-reddened color, minimum errors for
metallicity estimation were obtained with the use of no more than six nodes
arranged in a single layer.  The overall one-sigma error in prediction of
metallicity was $\sigma \feh = 0.26$ dex, with a median offset of
+0.04 dex (note that this prediction error {\it includes} the errors in the 
metallicities of the Beers et al.  1999 standards themselves).  Division of the
validation set into several intervals of color and (known) metallicity did not
reveal any large deviations from these error levels over the calibration space.
We list the resulting abundance estimates, [Fe/H]$_{\rm ANN1}$, in column (3)
of Table 6.  Inspection of the comparison between the two ``photometric''
abundance indicators reveals that agreement is generally excellent, and in most
cases, within the quoted one-sigma error estimate.  All of the derived
abundances agree within two sigma.  Since the majority of the error in the
``photometric'' abundance indicators probably arises from difficulties in the
proper estimation of the reddening correction, we explore alternative
approaches as described below.  Once near-IR $JHK$ photometry from the final
release of the 2MASS point source catalog (Skutskie et al. 1997) becomes
available, we will be able to predict de-reddened $(B-V)_{0}$ colors with more
confidence.

\subsubsection{Estimates Using Spectral Information Only}

We have trained yet another ANN, taking as inputs $LKP$, $LACF$, and $LDGP$,
and producing as output an estimate of the metallicity [Fe/H].  The training
was carried out using the subset of 398 of the 551 standard stars from 
Beers et al. (1999) for which measures of all three inputs were available,
setting aside 20\% of this sample for use as a validation set to estimate
errors in the procedure.  Minimum errors for metallicity estimation were
obtained with the use of no more than six nodes arranged in a single layer.
The overall one-sigma error in prediction of [Fe/H] was $\sigma \feh = 0.29$
dex, with a median offset of $-0.02$ dex.  Division of the validation set into
several intervals of $LDGP$ and (known) metallicity did not reveal any large
deviations from these error levels over the calibration space.  We list the
resulting abundance estimates, [Fe/H]$_{\rm ANN2}$, in column (4) of Table 6.

Snider et al. (2001) describe a procedure for the use of ANNs that take, as
inputs, the entire set of spectral information (after normalization of
the spectral energy distribution) over the (minimum) wavelength range
$3850-4450$ \AA\ , and produce as output an estimate of [Fe/H], with an overall
one-sigma scatter of about 0.20 dex.  We have attempted to make use of this
procedure for the present sample of stars, although we were somewhat hampered
by resolution limitations, as described below.

All spectra were first re-binned to the nominal dispersion of the trained ANNs
used by Snider et al. (2001), 0.65 \AA/pixel.  This was a relatively minor
change for the spectra obtained with the CTIO 4m and ESO 1.5m, but required a
rather severe over-sampling of the data obtained with the ESO 3.6m.  The
spectra were then submitted to the network described by Snider et al. as the
``total/full'' network, details of which can be found in their paper.  This
network is based on the subset of 279 stars from Beers et al. (1999) with
previously observed high S/N medium-resolution spectroscopy available, with a
lower S/N limit of about 40/1 (at the red end of the spectra). 

The estimated abundances which result from this approach, [Fe/H]$_{\rm ANN3}$,
are listed in column (5) of Table 6.  As can be seen from inspection of the
table, for the most part the resulting abundances are consistent, within 0.5
dex, with the estimated metallicities based on the other approaches we
have employed.  In a number of cases, however, the [Fe/H]$_{\rm ANN3}$ did not
agree very well.  We have indicated these cases in the tables by putting the
more doubtful results in parentheses.  The reasons for these disagreements may
involve a number of sources:  (1) Three of the spectra with gross deviations
are from the ESO 3.6m, which as we commented above, had to be over-sampled in
order to run them through the previously trained network, and (2) The network
used to evaluate our stars is not populated with large numbers of metal-poor
giants, and gaps in the coverage of the pertinent ranges of this parameter
space may be a limiting factor.  Despite these difficulties, the consistency
in metallicity estimates obtained for the majority of the program stars from
this method, as compared to the other approaches, provides confidence in this
technique.  It was suggested by an anonymous referee that we consider dropping
the [Fe/H]$_{\rm ANN3}$ estimates of abundances in our final averages.  We have
decided not to follow this advice (except for the problematic cases), on the
grounds that these estimates are based on a completely different (albeit new,
and less than optimally tested) calibration that, unlike all of our other
approaches, does not involve individual line index measurements.

\subsubsection{Final Adopted Metallicities and Comparison with Available High-Resolution Abundance Estimates}

We obtain our final abundance estimates from a straight average of the four
derived abundances for each star listed above -- two ``photometric'' and two
``non-photometric.''  In the case of the rejected [Fe/H]$_{\rm ANN3}$
estimates, we have simply dropped these from the averaging.  The final
estimates of metallicity, [Fe/H]$_{\rm F}$, are listed in column (6) of Table
6.  Although we do not have individual one-sigma error estimates for the
[Fe/H]$_{\rm ANN3}$ results, the Snider et al.  (2001) results lead us to
believe that they should be on the order of $\sigma \feh = 0.2$ dex, similar to
the errors we were able to obtain from the application of the Beers et al.
(1999) calibration.  Certainly the range of values reported in Table 6, from
the application of different abundance estimation procedures, supports this
assumption. A comparison of the average metallicity obtained from
the first three estimates listed in Table 6 ([Fe/H]$_{\rm AK2}$, [Fe/H]$_{\rm
ANN1}$, [Fe/H]$_{\rm ANN2}$) with the 33 accepted [Fe/H]$_{\rm ANN3}$ estimates
indicates the presence of a zero-point offset of only +0.03 dex, and a
one-sigma scatter of 0.31 dex of [Fe/H]$_{\rm ANN3}$ with respect to the other
methods, consistent with expectations.  

The use of multiple metallicity estimation procedures, relying on different
inputs (and different calibrations), will serve to decrease the systematic
errors associated with any single method.  Ultimately, the errors in our
determination of metallicity are driven by the accuracy of the abundances
assigned to the Beers et al. (1999) standards, so we conservatively adopt a
global (external) error estimate of 0.2 dex to our final abundance estimates.

Among the LSE metal-poor candidates we have re-discovered the bright
metal-poor giant HD 184711 (LSE-149), for which the average abundance
reported by Beers et al. (1999), based on high-resolution spectroscopic
measurements, is $\feh = -2.51$.  The agreement with the final abundance
reported in Table 6, $\feh_{\rm F} = -2.52 \pm 0.20$, is excellent.  Another of
our program stars, LSE-182, is the bright giant HD 178443, for which McWilliam
et al. (1995b) obtained an abundance estimate of $\feh = -2.07$.  This is
somewhat lower than we have assigned, $\feh_{\rm F} = -1.68 \pm 0.20$, but only
by about 2 sigma (disregarding the error in the high-resolution estimate).

Several LSE stars were targeted for high-resolution study as part of a
recently completed Large Programme with VLT/UVES by Cayrel et al..  These
included the most metal-deficient star in the sample, LSE-131 ($\feh_{\rm F} =
-2.62$), and two stars of somewhat higher abundance, but with kinematics (as
discussed below) that suggest possible association with the MWTD, LSE-173
($\feh_{\rm F} = -1.92$) and LSE-232 ($\feh_{\rm F} = -2.35$).  Final abundance
estimates from the Cayrel et al. UVES observations have not been obtained as of
yet, but preliminary inspection of the high-resolution spectrum for LSE-131
confirms that its abundance is consistent with $\feh \approx -2.5$, or slightly
lower.  A previous high-resolution spectrum of LSE-131, obtained with the ESO
3.6m telescope, and reported by Spite et al. (1999), suggests an abundance
$\feh = -2.8$, in close agreement with the estimated abundance obtained in the
present paper.

We conclude that our abundance estimates should be trusted, and we
proceed with our kinematic analysis below.

\section{{\it Hipparcos} and {\it Tycho-II} Proper Motions, and Derived Space Motions}

Ten stars in the present program were included in the {\it Hipparcos} catalog,
with average accuracies of 2.43 mas~yr$^{-1}$ in $\mu_{\alpha^\ast}$
($=\mu_\alpha \cos\delta$) and 1.86 mas~yr$^{-1}$ in $\mu_\delta$,
respectively.  Columns (6)--(9) of Table 7 list $\mu_{\alpha^\ast}$,
$\mu_\delta$, and their associated errors, as given in the {\it Hipparcos}
catalog.  All of these same stars, as well as the fainter ones, have proper
motions available from the {\it Tycho-II} catalog, with average accuracies of
2.06 mas~yr$^{-1}$ in $\mu_{\alpha^\ast}$ and 1.99 mas~yr$^{-1}$ in
$\mu_\delta$, respectively.  Columns (2)--(5) of Table 7 list the proper
motions and their associated errors as given in \ the {\it Tycho-II} catalog.
As in Beers et al. (2000), we construct a variance-weighted average of the
available proper motions.  These averages, and their associated errors, are
listed in columns (10)--(13) of Table 7, respectively.

\subsection{Space Motions for the LSE Stars}

We now derive the space motions and orbital parameters of the LSE stars,
following the procedures of Beers et al. (2000); Table 8 provides a summary
of the results.  Column (1) lists the star name.  Column (2) recalls
the derived metallicity from Table 6.  Columns (3) and (4) list the positions
of the stars in the meridional plane $(R,Z)$, adopting $R_\odot = 8.5$ kpc as
the Galactocentric distance for the Sun.  Columns (5)--(7) list the
three-dimensional velocities $U$, $V$, and $W$, in the directions toward the
Galactic anticenter, the rotational direction, and the north Galactic pole,
respectively, along with an estimate of the errors in these quantities that
could arise from errors in distance estimates of 20\%, as described below.
These velocity components are corrected for the solar motion
$(U_\odot,V_\odot,W_\odot)=(-9,12,7)$ km s$^{-1}$ with respect to the local
standard of rest (LSR) (Mihalas \& Binney 1981).  Columns (8) and (9) list the
velocity components $(V_R,V_\phi)$ in the cylindrical rest frame $(R,\phi)$,
respectively, on the assumption that the rotational speed of the LSR around the
Galactic center is $V_{LSR} = 220$ km s$^{-1}$.

To estimate the orbital parameters for these stars, we adopt the analytic
St\"ackel-type mass model developed by Sommer-Larsen \& Zhen (1990), which
consists of a flattened, oblate disk and a nearly spherical massive halo. This
model reproduces a flat rotation curve beyond $R=4$ kpc and the local mass
density at $R_\odot$, consistent with other observations.  Columns (10) and
(11) of Table 8 list the estimated apogalactic distances, $R_{ap}$, and the
estimated perigalactic distances, $R_{pr}$, along the Galactic plane,
respectively.  Column (12) lists the maximum distance above (or below) the
plane, $Z_{max}$, explored by each star in the course of its orbital motion.
In column (13) we list the characteristic eccentricities of the orbits, defined
as $e = (r_{ap}-r_{pr})/(r_{ap}+r_{pr})$, where $r_{ap}$ and $r_{pr}$ stand for
the apogalactic and perigalactic distances from the Galactic center,
respectively.

An anonymous referee suggested that we investigate the impact of possible
distance errors on our derived kinematic quantities.  We carried out this
exercise by repeatedly subsampling from our catalog of program stars, with the
listed distances of the stars perturbed by 10\%, 20\%, and 30\%, respectively,
then re-deriving the quantities $UVW$ and $e$ within our adopted potential.
For completeness, we also included the effects of an assumed radial velocity
errors of 10 km s$^{-1}$, and the listed errors in the adopted proper motions.
The average errors, for the entire set of program stars, obtained from this
procedure were as follows:

\centerline{10\% errors in distance:  $<\epsilon(U,V,W)>\; =  (~8, ~8,~5) $ km s$^{-1}$}
\centerline{20\% errors in distance:  $<\epsilon(U,V,W)>\; =  (10, 11,~7) $ km s$^{-1}$}
\centerline{30\% errors in distance:  $<\epsilon(U,V,W)>\; =  (12, 14,~8) $ km s$^{-1}$}
\bigskip
\centerline{10\%, 20\%, 30\% errors in distance:  $<\epsilon(e)>\; = $ 0.03, 0.04, 0.05}

Table 8 includes values of the expected errors in the kinematic quantities for
individual stars arising from assumed 20\% errors in the distance estimates.
Note that in all but one case (the most distant star, LSE-118), the likely
errors in the derived kinematic quantities are quite small, thus are not
expected to significantly affect the interpretation of our results.

\subsection{The Local Fraction of Metal-Weak Thick Disk Stars}

As noted in the Introduction, previous (non-kinematically biased) searches for
metal-deficient stars have concentrated primarily on high Galactic latitudes
(the notable exception being MFF, where the existence of the MWTD was first
suggested).  This surely has introduced an underestimate of the numbers of
nearby MWTD stars, so we were curious to compare the relative fractions of
likely MWTD stars in the LSE survey with previous work.  As a representative
comparison sample, we have selected the 412 giants with $V < 12.0$ and [Fe/H]
$\le -0.6$ from the Beers et al. (2000) catalog with available space motions
and orbital eccentricities from Chiba \& Beers (2000).  An anonymous referee
pointed out that, by selecting stars from this catalog with available space
motions, one runs the risk of unintentionally re-introducing kinematic biases
into our comparison sample.  Although this certainly is a concern, the original
non-kinematical selection of stars in the Beers et al. (2000) sample, from
which the Chiba \& Beers (2000) catalog was drawn, should minimize this
problem.  In any event, the inhomogeneous nature of the sample assemblage
precludes the possibility of making explicit corrections for possible biases, a
fact that should be kept in mind by the reader.

Figure 2 (panels a-c) is a plot of the $U,V,W$ velocity components for the 36
LSE giants and subgiants (solid circles) of the present investigation, as well
as for the three stars we classify as field horizontal-branch (FHB) or
main-sequence turnoff (TO) stars (open circles).  For the purpose of the
kinematic analysis we have eliminated the one star classified as FHB in Table
5, as well as the two stars classified as TO.  Figure 2 (panels d-f) shows the
same information for the comparison sample described above.  It is immediately
clear that many of the LSE giants exhibit rather small $V$ velocities,
suggesting possible membership in a rapidly rotating population, and small $W$
velocities, suggesting that they are drawn from a population with low vertical
velocity dispersion.  The distribution of $U$ velocities exhibits a rather
higher dispersion.  This characteristic has been noted in previous samples, but
its origin has not yet been satisfactorily explained in the context of present
models of Galactic structure, even after attempts to account for
selection-related biases (see the discussion of samples considered by Ryan
\& Norris 1991).  The comparison sample of bright giants includes a large
number of stars that are clear members of the halo population, as may be
inferred from the relatively broad distribution of the individual velocity
components below $\feh = -1.5$.

The derived mean velocities and velocity ellipsoids of the LSE sample and the
comparison sample are summarized in Table 9.  Although the small numbers of
stars limits the accuracy with which the ellipsoid for the LSE stars can be
determined, close inspection of these results reveals a few interesting
differences between the two samples.  First, note that for the comparison
sample, $<V>$ changes dramatically, from a moderate velocity lag on the order
of $-60$ \kms\ in the metallicity range $-1.6 \le\; {\rm [Fe/H]}\; \le -0.6$, to
a velocity lag of roughly $-180$ \kms\ for metallicies below [Fe/H] = $-1.6$.
In contrast, the LSE sample exhibits a velocity lag that remains essentially
constant, centered around $<V> = -80\; \kms$ over the different cuts in
metallicity.  This strongly indicates that the kinematics of the population(s)
of stars that the LSE sample are drawn from are rather different from those
that are sampled by the comparison sample.  Furthermore, note that at the
lowest metallicity cutoff, two of the three components of the LSE sample
velocity ellipsoid ($\sigma_V$ and $\sigma_W$) appear significantly lower than
the corresponding components of the comparison sample.  Interestingly, in the
metallicity range $-1.6 \le\; {\rm [Fe/H]}\; \le -0.6$, the $\sigma_U$ component
of the LSE star velocity ellipsoid appears marginally {\it greater} than the
corresponding component of the comparison sample.  Again, these results
suggests the lack of a common parent population.

The differences between the populations highlighted above can be shown most
clearly by contrasting the distribution of $V_{\phi}$ for the LSE giants
with that of the comparison sample.  Figure 3a shows a stripe
density plot of $V_{\phi}$ for the 34 LSE giants with [Fe/H] $\le -1.0$ (all of
which have $|Z| \le 1$ kpc).  Figure 3b shows the same diagram for the subset
of the 164 giants in the comparison sample with $\feh \le -1.0$ and $|Z| \le 1$
kpc.  Note that, based on the previous discussion of Chiba \& Beers 2000, we
expect that the comparison sample in this metallicity range may indeed contain
a significant number of MWTD stars; these can be seen in Figure 3b as the
concentration of lines in the broad velocity interval $150 \le V_{\phi} \le
250$ \kms. Of course, this same velocity interval will contain numerous members
of the halo population as well, due to its large velocity dispersion.  Note,
however, that the comparison sample {\it also} contains a large number of stars
with velocities we would uniquely associate with the halo population, i.e.,
$V_{\phi} < 100$ \kms.  Inspection of Figure 3a suggests that, while the LSE
sample certainly contains a handful of halo objects, the concentration of lines
in the interval $150 \le V_{\phi} \le 250$ \kms\ is more pronounced than seen
in the comparison sample.  A two-sample K-S test supports these impressions.
The hypothesis that the two samples are drawn from a common parent is rejected
with probability $p = 0.042$ (two-sided).  A one-sided test, where the
alternative hypothesis is that the LSE stars are drawn from a population of
higher mean rotation, is of course an even stronger rejection.

Figures 3c and 3d show similar plots as described above, but for the
metallicity cut $\feh \le -1.6$, the metallicity below which most previous
authors have argued that the MWTD ceases to make an important contribution to
the local volume density of metal-poor stars.  Note that while the distribution
of the 100 stars in the comparison sample seen in Figure 3d is broad and
roughly symmetric about $V_{\phi} \approx 50\; \kms$, consistent with its being
composed primarily of halo objects, the distribution of the 24 LSE stars in
Figure 3c is clearly centered on much higher rotational velocities; in fact the
lower cut on metallicity has removed most of the LSE stars we might have
associated with the halo population!  Not surprisingly, a K-S test rejects the
likelihood of these samples sharing a common parent at a very high level, $p =
0.006$ (two-sided).

One might wonder whether some selection bias has produced the rather different
distributions of $V_{\phi}$ described above.  After all, the comparison sample
was drawn from numerous samples covering much of the high-Galactic latitude
sky, while the LSE sample came from a more limited range in Galactic longitude
($-60\deg \le l \le +60\deg$) at lower latitudes.  In fact, the selection is
rather stronger than this, as absorption toward the Galactic center has
eliminated most of the sample within thirty degrees of $l=0\deg$, as can be
seen from inspection of Table 1.  To assess whether the different longitude
selections have conspired to produce the rather different $V_{\phi}$
distributions, Figure 4 shows similar diagrams as in Figure 3, but with the LSE
longitude cuts included in the sub-selection of the comparison sample.
Although there are of course fewer stars in the comparison sample after these
restrictions, the visual impression of the difference in the distributions
remains.  A two-sample K-S test of the subsamples of 34 LSE stars and 103
comparison-sample stars with [Fe/H] $\le -1.0$, shown in Figures 4a and 4b,
respectively, rejects the common parent hypothesis at a high level, $p = 0.002$
(two-sided).  For the 24 LSE stars and 67 comparison-sample stars with [Fe/H]
$\le -1.6$, shown in Figures 4c and 4d, respectively, the rejection is even
stronger, $p = 0.001$ (two-sided).

Figure 5(a) shows the relation between $e$ and [Fe/H] for the LSE stars.  There
clearly exists a non-negligible fraction of low-eccentricity metal-poor stars
in this sample (again, the three non-giants are shown with open circles).  Over
60\% (22 of 36) of the LSE giants exhibit eccentricities less than $e = 0.5$.
Figure 5(b) shows these same quantities for the comparison sample.  In this
panel, the filled circles represent the stars in the Galactic longitude range
$-60\deg \le l \le +60\deg$, while the open circles represent the stars outside
of this range.  The visual impression one obtains is that the numbers of stars
at low metallicity and low eccentricity in the comparison sample has been {\it
decreased} by the application of the cuts in Galactic lontitude that are
pertinent to the LSE sample.  This runs counter to the notion that the
longitude selection of the LSE sample has somehow overemphasized the importance
of the low eccentricity stars.  In fact, one might be tempted to conclude that
more complete longitude coverage at low latitudes would be likely to boost the
relative numbers of low metallicity, low eccentricity stars.

For a more quantitative comparison, we show in Figure 6a the cumulative $e$
distributions, $N(<e)$, in the abundance ranges [Fe/H] $\le-1.0$ (thin dashed
histogram) and [Fe/H] $\le-1.6$ (thin solid histogram) for the 36 LSE giants,
all of which have $|Z|<1$ kpc.  In this same panel we also plot $N(<e)$ for the
comparison sample of bright giants with $|Z|<1$ kpc. Inspection of this figure
suggests that the LSE sample contains more nearly circular orbits at [Fe/H]
$\le-1.6$ than the comparison sample (thick solid histogram), whereas at [Fe/H]
$\le-1.0$ (the thick dashed histogram representing the comparison sample), the
difference, if any, in $N(<e)$ is less clear. An anonymous referee pointed out
that it appeared from inspection of Figure 5 that the ``halo objects,'' which
one might loosely define to be those with eccentricities exceeding $e = 0.5$,
appeared to have somewhat lower metallicities than expected if fair draws were
made from the halo population.  This effect, if real (small number statistics
prevent any solid judgement to be made), is surely driven by the original
selection of the LSE giants as metal-poor candidates.  In any event, the same
selection criteria were used for all of the candidate stars prior to any
knowledge of their kinematics, hence the differential comparisons we have
carried out are still meaningful.

A two-sample K-S test indicates that the eccentricity distributions for the
cut in metallicity [Fe/H] $\le -1.0$ cannot be distinguished; rejection of the
hypothesis that the subsamples are drawn from the same parent population is not
significant ($p = 0.25$, one-sided, versus the alternative that the LSE stars
are drawn from a parent population with lower eccentricity).  However, for the
cut in metallicity [Fe/H] $\le -1.6$, a K-S test is able to reject the common
parent population hypothesis at a marginally significant level, $p = 0.055$
(one-sided).  The ``near rejection'' of the subsample of stars with [Fe/H] $=
-1.6$ is certainly suggestive, though not yet definitive.  Interestingly, when
we apply the longitude cuts to the comparison subsample, as discussed above, in
order to make it match the longitude distribution of the LSE subsample, it is
possible to significantly reject the common parent hypothesis for both of the
metallicity cuts; for [Fe/H] $\le -1.0$, $p = 0.022$ (one-sided), for [Fe/H]
$\le -1.6$, $p = 0.009$ (one-sided).

The above analysis certainly indicates a clearer signature of the MWTD
population if the sample is selected at low Galactic latitude, as in the
present work. To confirm this, we estimate the contribution of the
thick-disk component, $F_{\rm MWTD}$, amongst local samples of metal-poor
stars, using the derived distribution of $e$.  Following the method developed
by Chiba \& Yoshii (1998), we have performed a Monte Carlo simulation to
predict the $e$ distribution from a mixture of stars contributed by the
thick-disk and halo populations. The characteristic kinematic parameters for
these components are drawn from Chiba \& Beers (2000): $<V_\phi>=33$
km~s$^{-1}$ and $(\sigma_U,\sigma_V,\sigma_W)=(141,106,94)$ km~s$^{-1}$ for the
halo, and $<V_\phi>=200$ km~s$^{-1}$ and
$(\sigma_U,\sigma_V,\sigma_W)=(46,50,35)$ km~s$^{-1}$ for the thick disk.
Figure 4b (thick solid line) shows the results of this exercise for $F_{\rm
MWTD}=0$, $F_{\rm MWTD}=0.3$, and $F_{\rm MWTD}=0.4$. It is evident that the
eccentricity distribution of the LSE giants with [Fe/H]  $\le-1.6$ is
characterized by $F_{\rm MWTD} \sim 0.3$, substantially larger than the
estimate of $F_{\rm MWTD} \sim 0.1$ derived from the sample considered by Chiba
\& Beers (2000).  With the metallicity cut [Fe/H]  $\le -1.0$, the value of
$F_{\rm MWTD}=0.4$ appears to be a superior fit.  {\it Both results strongly
suggest that previous non-kinematic selection of metal-poor stars at higher
Galactic latitudes has resulted in a severe underestimate of the relative
importance of the MWTD in local samples.}

\subsection{Assignment of Population Membership}

As seen from the discussion above, many of the LSE stars exhibit rather small
$V$ velocities, suggesting that they may belong to a rapidly rotating (thick)
disk component; we now attempt to assign the likely population membership of
each LSE star based on its full space motion.  This is clearly an inexact
procedure, since the halo population exhibits large dispersions in all of its
velocity components.  If the motion of a star is well outside an acceptable
range of the characteristic spatial and velocity distributions of the thick
disk, it is most likely a member of the halo population, otherwise it belongs
to {\it either} the disk or halo population, and we cannot uniquely
determine its membership.

The velocity distribution of the thick-disk component was determined by Chiba
\& Beers (2000), using a large number of stars from the Beers et al. (2000)
catalog, summarized as $<V_\phi>_{disk}=200$ km~s$^{-1}$ and
$(\sigma_{U,disk},\sigma_{V,disk},\sigma_{W,disk})=(46,50,35)$ km~s$^{-1}$.  We
also adopt $|Z|\le 1$ kpc as a typical vertical range of the thick disk (Chiba
\& Yoshii 1998; Chiba \& Beers 2000). If a star exhibits $|Z_{max}|>1$ kpc,
or at least one of its velocity components deviates from the above velocity
range of the disk at more than a 2-$\sigma$ level, we assign it to the halo
population, denoted as ``H'' in column (14) of Table 8. On the other hand, a
star within the above range of the disk at less than a 2-$\sigma$ level might
belong to either the disk or halo population, which we label as ``DH'' in
column (14).  The three stars with metallicities $\feh > -0.50$ also exhibit
space motions expected for membership in a disk population, hence we assign the
classification ``D'' in column (14). 

Since there is great interest in searches for any chemical signature of the
origin of the MWTD, we have noted with asterisks the stars in Table 8 that are
classified as ``DH,'' but having low (absolute values of) individual velocity
components (taken here to mean $|UVW| \le 50$ \kms), and that further satisfy
the requirements $V_{\phi} \ge\;  170\; \kms$, $Z_{max} < 1$ kpc, $\feh =
-0.6$, suggesting that they may indeed be bona-fide members of the MWTD, and
hence deserving of detailed study at high-resolution.  This sample may not be
pure, but it seems likely that at least a number of these stars are members of
the MWTD population.  Note that the familiar metal-poor giant HD~184711 (=
LSE-149) just misses designation as a likely member of the MWTD, since its $V$
velocity component is somewhat higher than the above criteria allow.

\section{Summary and Discussion}

We have presented spectroscopy and photometry for a small sample of bright
metal-deficient giant candidates selected from a prism survey (the LSE survey
of Drilling \& Bergeron 1995) that explores lower Galactic latitudes than most
previous surveys for metal-deficient stars.  Estimates of metallicity for the
stars in this sample have been obtained by a variety of methods, all in good
agreement with one another.  Since all of our program stars have available
proper motions, we were able to derive estimates of their complete space
motions and orbital eccentricities. 

Inspection of the distribution of rotational velocities for the LSE stars
indicates that they cannot be drawn from the same parent population as stars
from previous samples of similarly bright giants (generally selected at higher
Galactic latitude), such as described by Beers et al. (2000); many individual
stars appear to be rotating quite rapidly about the Galactic center.
Furthermore, inspection of the distribution of orbital eccentricity for the LSE
giants, as contrasted with that of the same comparison sample of bright giants,
has revealed that the LSE sample contains a much larger proportion of
metal-weak stars with low eccentricities, as might be expected if the MWTD
population is an important component in the solar neighborhood.  Our best
estimates of the fraction of local MWTD stars, based on Monte Carlo models of
the expected distribution of orbital eccentricities of a pure halo population,
suggest $F_{\rm MWTD} \approx 40\%$ for the metallicity regime $\feh \le -1.0$,
and remaining as high as $F_{\rm MWTD} \approx 30\%$ for the metallicity regime
$\feh \le -1.6$.  This fraction is {\it triple} the value obtained for stars
with $\feh \le -1.6$ in the Chiba \& Beers (2000) analysis of the stars in the
Beers et al. (2000) catalog.  The lowest metallicity star in the LSE sample
with kinematics that are consistent with membership of the MWTD population is
LSE-156, with [Fe/H]$ = -2.35$.

Over the past decade, a number of claims for a significant population of
metal-poor stars with disk-like kinematics have been made, but acceptance of
their presence has been cast in doubt because of incorrectly assigned 
metallicities.  Based on this new sample, this no longer appears to be the
case, and we must endeavor to understand the implications of a significant
population of MWTD stars for theories of the formation and evolution of the
Galaxy.  In this respect, it is important to keep in mind that, although the
MWTD population may contribute a large fraction of the {\it local} metal-poor
stars, the (inner) halo population is probably still the dominant reservoir of
stars with [Fe/H] $ \le -1.6$ within a few kpc of the Sun.  Furthermore,
although we have emphasized the possible importance of the MWTD population, it
certainly appears to be a minor constituent of the entire thick-disk
population;  Martin \& Morrison (1998) suggest that the local density of the
MWTD represents less than 1\% of that of the canonical thick disk.

It is of interest to note that the comparison of [Fe/H] versus orbital
eccentricity diagrams of Chiba \& Beers (2000) with the numerical models of
hierarchical galaxy formation of Bekki \& Chiba (2001) suggested that the
models were {\it overproducing} the expected numbers of metal-poor stars with
low eccentricities relative to the observations (see Figure 14 of Bekki \&
Chiba 2001), at least in the intermediate abundance range $-1.6 \le \feh \le
-1.0$.  It now seems likely that the problem may lie, at least in part, with
the observations themselves, which have not extended to sufficiently low
Galactic latitudes to fairly sample the presence of MWTD stars.

If, as we have argued, their does indeed exist a significant fraction of 
thick-disk stars with metal abundances [Fe/H]$\le -1.6$, this finding may have
significance to formation scenarios for the Milky Way, and by inference, for
other large spiral galaxies.  One presently plausible explanation for the
origin of a MWTD component may be the merging of small proto-Galactic
fragments (e.g., Searle \& Zinn 1978) with a pre-existing thin, possibly
metal-poor stellar disk (e.g., Quinn, Hernquist, \& Fullagar 1993; Wyse 2001).
Such fragments may correspond to the progenitors of the present-day luminous
dwarf satellites, such as Sagittarius (Ibata, Gilmore, \& Irwin 1994), or some
of the numerous cold-dark-matter subhalos surrounding the Galaxy, as predicted
from recent cosmological simulations (e.g., Klypin et al. 1999; Moore et al.
1999).  Minor merging events might also explain the origin of the rapidly
rotating, thick-disk globular clusters (Bekki \& Chiba 2002).  Recent
identification of various stream-like features in the halo (and possibly near
the disk) may be associated with the debris of these past merging events (Wyse
et al. 2000; Newberg et al. 2002).  Dinescu (2002) has argued, from a close
inspection of the Beers et al. (2000) sample, for the presence of a retrograde
population that exhibits similarities to the orbit of the globular cluster
$\omega$-Centauri.  Derivation of a more precise estimate of the fractional
contribution of the MWTD component in the solar neighborhood will help set
limits on the merging process(es) in the early (and possibly more recent)
Galaxy.

One key piece of information for the likely source of the MWTD stars is
obtainable by study of the relative abundance patterns of individual elements
for stars of the thick-disk population.  Recently, Prochaska et al. (2000) have
carried out such a study, based on ten stars with disk-like kinematics chosen
from the proper-motion selected survey of Carney et al. (1994), covering
the metallicity range $-1.0 \le \feh \le -0.4$, the range most pertinent to the
canonical thick disk.  These authors concluded that the thick-disk elemental
abundance patterns were essentially identical to those for stars of the halo
population, consistent with the idea that the two populations share similar
nucleosynthesis histories.  It is of obvious importance to extend such studies
to lower metallicities, such as could be accomplished by abundance analyses of
the LSE stars noted in the present paper, to see if this result applies to
stars in the abundance range $-2.5 \le \feh \le -1.0$.  Another useful set of
targets for high-resolution studies may be found in Table 6 of Chiba \& Beers
(2000).  This last point is crucial, as previous studies of Galactic
chemical evolution have generally adopted the view that stars with
metallicities below [Fe/H] $\approx -1$ represent an essentially pure halo
population.  Unless caution is taken (for example, by only using those stars
with inferred distances more than a few kpc above the disk plane, or with
kinematics that are indisputably associated with the halo), there is the clear
danger of confounding the sample with mixed populations.

Clearly, it would also be important to carry out further surveys for the
detection of bright (hence nearby) metal-poor stars at lower Galactic
latitudes.  One attractive sample could be assembled from the extensive
re-classifications of the HD catalog stars by Houk et al. (Houk \& Swift 1999,
and references therein). Inspection of the available data reveals that there
are several hundred bright F- and G-type stars, classified as possibly
metal-deficient, located at Galactic latitudes $|b| \le 30\deg$, many of which
already have available proper motions.  A medium-resolution spectroscopic
survey of these stars is just now getting underway, and should provide
important constraints on the MWTD population in the near future.

\acknowledgments

TCB acknowledges partial support for this work from grants AST 95-29454, AST
00-98549 and AST 00-98508 from the National Science Foundation.  MC
acknowledges partial support from Grants-in-Aid for Scientific Research
(09640328) of the Ministry of Education, Science, Sports and Culture of Japan.
SR acknowledges partial support for this work from grant 200068/95-4 CNPq,
Brazil, and from the Brazilian Agency FAPESP.  TvH acknowledges partial support
from grant AST 98-19768 from the National Science Foundation.  We thank
Johannes Reetz for obtaining several spectra for us at the ESO 3.6m telescope.
We would also like to thank the anonymous referee, whose comments improved the
presentation of our results.

This work made use of the SIMBAD database, operated at CDS, Strasbourg, France.

\clearpage

\figcaption[fig1.eps]{Example spectra of four LSE giants with similar
de-reddened colors, and with metallicities obtained as described in the text,
arranged from relatively metal-rich to relatively metal-poor.  The spectra have
been normalized to a continuum approximately equal to unity, and shifted to
zero rest velocity.  Note that the original spectra extended redder than shown;
the region depicted in the figure is meant to emphasize the metallic features
that drive the metallicity estimates. \label{fig1}}
 
\figcaption[fig2.eps]{(panels a-c) Local velocity components, $U,V,W$, for the
LSE giants, and (panels d-f) for a sample of bright giants with $V \le 12$
and [Fe/H] $\le -0.6$ from the Beers et al. (2000) catalog.  The three points
depicted with open circles in panels (a-c) are classified as either TO or FHB.
\label{fig2}}

\figcaption[fig3.eps]{Stripe density plots of the derived rotational velocity
component $V_{\phi}$ for (a) LSE giants with [Fe/H] $\le -1.0$, (b) Giants in
the comparison sample with [Fe/H] $\le -1.0$, (c) LSE giants with [Fe/H] $\le
-1.6$, and (d) Giants in the comparison sample with [Fe/H] $\le -1.6$.
Note that the LSE giants exhibit a higher fraction of stars with
large (positive) $V_{\phi}$ than the comparison sample, for both metallicity
cuts. \label{fig3}}

\figcaption[fig4.eps]{Stripe density plots of the derived rotational velocity
component $V_{\phi}$ for (a) LSE giants with [Fe/H] $\le -1.0$, (b) Giants in
the comparison sample with [Fe/H] $\le -1.0$ and selected in a longitude
interval similar to the LSE giants, (c) LSE giants with [Fe/H] $\le -1.6$, and
(d) Giants in the comparison sample with [Fe/H] $\le -1.6$ and selected in a
longitude interval similar to the LSE giants.  The distributions still appear
quite different from one another, at both metallicity cuts. \label{fig4}}

\figcaption[fig5.eps]{(a) Distribution of [Fe/H] for the LSE giants as a function
of derived orbital eccentricity.  Note the presence of substantial numbers of
stars with quite low metallicity even for $e \le 0.5$.  Open circles indicate
the non-giants. (b) The same as in panel (a), but for the giants in the
comparison sample.  In this panel, the filled circles represent stars chosen to
satisfy $-60\deg \le l \le +60\deg$, while the open circles represent stars
outside of this longitude range. \label{fig5}}

\figcaption[fig6.eps]{(a) Cumulative eccentricity distributions of the LSE
giants for metallicity cuts of [Fe/H]$\le-1.0$ (thin dashed histogram) and
[Fe/H]$\le-1.6$ (thin solid histogram). The thick dashed and solid histograms
denote the comparison sample of giant stars at $|Z|<1$ kpc in these same
abundance ranges.  (b) Comparison of the cumulative eccentricity distributions
of the LSE giants with Monte Carlo models, based on a mixture of two Gaussian
components taken to represent the halo and thick disk, where the disk fraction
is denoted as $F$.  We take $<V_\phi>=33$ km~s$^{-1}$ and
$(\sigma_U,\sigma_V,\sigma_W)=(141,106,94)$ km~s$^{-1}$ for the halo, and
$<V_\phi>=200$ km~s$^{-1}$ and $(\sigma_U,\sigma_V,\sigma_W)=(46,50,35)$
km~s$^{-1}$ for the thick disk.
\label{fig6}}


\begin{thebibliography}{}

\bibitem[]{} Arce, H.G., \& Goodman, A.A. 1999, \apj, 512, L135

\bibitem[]{} Beers, T.C., \& Sommer-Larsen, J. 1995, \apjs, 96, 175

\bibitem[]{} Beers, T.C., Rossi, S., Norris, J.E., Ryan, S.G.,
\& Shefler, T. 1999, \aj, 117, 981.

\bibitem[]{} Beers, T.C., Chiba, M., Yoshii, Y., Platais, I., Hanson, R.B.,
Fuchs, B., \& Rossi, S. 2000, \aj, 119, 2866

\bibitem[]{} Bekki, K. \& Chiba, M. 2001, \apj, 558, 666

\bibitem[]{} Bekki, K., \& Chiba, M. 2002, \apj, 566, 245

\bibitem[]{} Bond, H.E. 1980, \apjs, 44, 517 

\bibitem[]{} Bonifacio, P., Centurion, M., \& Molaro, P. 1999, \mnras, 309, 533

\bibitem[]{} Burris, D.L., Pilachowski, C.A., Armandroff, T.E., Sneden, C.,
Cowan, J.J., \& Roe, H. 2000, \apj, 544, 302

\bibitem[]{} Burstein, D., \& Heiles, C. 1982, \aj, 87, 1165

\bibitem[]{} Carney, B.W., Latham, D.W., Laird, J.B., \& Aguilar, A. 1994, \aj,
107, 2240

\bibitem[]{} Chiba, M., \& Beers, T.C. 2000, \aj, 119, 2843

\bibitem[]{} Chiba, M., \& Yoshii, Y. 1998, \aj, 115, 168

\bibitem[]{} Chiba, M., Yoshii, Y., \& Beers, T.C. 1999, in The Third Stromlo
Symposium: The Galactic Halo, eds. B.K. Gibson, T.S. Axelrod, \& M.E. Putman
(San Francisco: ASP), 165, p. 273

\bibitem[]{} Dinescu, D.I. 2002, in Omega Centauri: A Unique Window into
Astrophysics, eds. F. van Leeuwen, G. Piotto, \& J. Hughes (San Francisco:
ASP), in press (astro-ph/0112364)

\bibitem[]{} Drilling, J.S., \& Bergeron, L.E. 1995, \pasp, 107, 846

\bibitem[]{} ESA 1997, The Hipparcos and Tycho Catalogues (ESA SP-1200)
(Noordwijk: ESA)

\bibitem[]{} Fuhrmann, K. 1998, \aap, 338, 161

\bibitem[]{} Hog, E., Fabricius, C., Makarov, V.V., Urban, S., Corbin, T.,
Wycoff, G., Bastian, U., Schwekendiek, P., \& Wicenec, A. 2000, \aap, 355, L27.

\bibitem[]{} Houk, N., \& Swift, C. 1999, Michigan Catalog of HD Stars, Vol. 5,
(Ann Arbor: Univ. of Michigan)

\bibitem[]{}Ibata, R., Gilmore, G. F., \& Irwin, M. J. 1994, \nat, 370, 194

\bibitem[]{} Johnson, H.L. 1963, in Basic Astronomical Data, ed. K. Aa. Strand
(Chicago: Univ. of Chicago Press), p. 204

\bibitem[]{} Johnson, H.L., Mitchell, R.I.,
Iriarte, B., \& Wisniewski, W.Z. 1966, Comm. Lunar Planet. Lab. 4, 99

\bibitem[]{} Katz, D., Cayrel, R., Coupry, M.-F., Perrin, M.-N., Van't Veer, C., Soubiran,
C., Barbuy, B., Bienayme, O., \& Friel, E. 1999, Astr. \& Space Science, 265,
p. 221

\bibitem[]{} Klypin, A., Kravtsov, A. V., Valenzuela, O., \& Prada, F. 1999, \apj, 522, 82

\bibitem[]{} Landolt, A.U. 1973, \aj, 78, 959

\bibitem[]{} Layden, A. 1995, \aj, 110, 2288

\bibitem[]{} Martin, J.C., \& Morrison, H.L. 1998, \aj, 116, 1724

\bibitem[]{} Mashonkina, L., \& Gehren, T. 2000, \aap, 364, 249

\bibitem[]{} McWilliam, A., Preston, G.W., Sneden, C., \& Shectman, S. 1995a, \aj, 109, 2736

\bibitem[]{} McWilliam, A., Preston, G.W., Sneden, C., \& Searle, L. 1995b, \aj, 109, 2757

\bibitem[]{} Mihalas, D., \& Binney, J. 1981, Galactic Astronomy (2d ed.;
San Francisco: Freeman)

\bibitem[]{} Moore, B., Ghigna, S., Governato, F., Lake, G., Quinn, T., \&
Stadel, J. 1999, \apj, 524, L19

\bibitem[]{} Morrison, H.L. 1993, \aj, 105, 539

\bibitem[]{} Morrison, H.L., Flynn, C., \& Freeman, K.C. 1990,
\aj, 100, 1191 (MFF)

\bibitem[]{} Morrison, J.E., R\"oser, S., Lasker, B.M., Smart, R.L., \& 
Taff, L.G. 1996, \aj, 111, 1405

\bibitem[]{} Newberg, H.J., Yanny, B., Rockosi, C.M., Grebel, E.K., Rix, H.-W.,
Brinkman, J., Csabai, I., Hennesey, G., Hindsley, R.B., Ibata, R., Ivezic, Z.,
Lamb, D., Nash, E.T., Odenkirchen, M., Rave, H.A., Schneider, D.P., Smith,
J.A., Stolte, A., \& York, D. 2002, \apj, in press

\bibitem[]{} Ryan, S.G., \& Norris, J.E., 1991, \aj, 101, 1835

\bibitem[]{} Norris, J., Bessell, M.S., \& Pickles, A.J. 1985, \apjs, 58 463

\bibitem[]{} Norris, J.E., Ryan, S.G., \& Beers, T.C. 2001, \apj, 561, 1034

\bibitem[]{} Prochaska, J.X., Naumov, S.O., Carney, B.W., McWilliam, A., \&
Wolfe, A.M. 2000, \aj, 120, 2513

\bibitem[]{} Quinn, P. J., Hernquist, L., \& Fullagar, D. P. 1993, \apj, 403, 74

\bibitem[]{} Ryan, S.G., \& Lambert, D. 1995, \aj, 109, 2068

\bibitem[]{} Schlegel, D.J., Finkbeiner, D.P., \& Davis, M. 1998,
\apj, 500, 525

\bibitem[]{} Schulte, D.H., \& Crawford, D.L. 1961, Kitt Peak Nat. Obs. Contrib. No. 10

\bibitem[]{} Searle, L., \& Zinn, R. 1978, \apj, 225, 357

\bibitem[]{} Skrutskie, M.F., Schneider, S.E., Stiening, R., Strom, S.E.,
Wienberg, M.D., Beichman, C., Chester, T. et al. 1997, in The Impact of Large
Scale Near-IR Sky Surveys, eds. F. Garzon et al. (Dordrecht: Kluwer), p. 187

\bibitem[]{} Snider, S., Allende-Prieto, C., von Hippel, T., Beers, T.C.,
Sneden, C., Qu, Y., \& Rossi, S. 2001, \apj, 562, 528

\bibitem[]{} Sommer-Larsen, J., \& Zhen, C. 1990, \mnras, 242, 10

\bibitem[]{} Spite, M., Spite, F., Cayrel, R., Hill, V., N\"ordstrom, B.,
Barbuy, B., Beers. T.C., \& Nissen, P.E., Astroph. \& Space Science 265, 141

\bibitem[]{} Stephenson, C.B., \& Sanduleak, N. 1971, Luminous Stars in the
Southern Milky Way, Pub. of the Warner and Swasey Obs., 1, 1

\bibitem[]{} Twarog, B.A., \& Anthony-Twarog, B.J. 1994, \aj, 107, 1371

\bibitem[]{} Wyse, R.F.G. 2001, in Galaxy Disks and Disk Galaxies,
eds. J.G. Funes \& E.M. Corsini (San Francisco:  ASP), 230, p. 71

\bibitem[]{} Wyse, R.F.G., Gilmore, G., Norris, J.E., \& Freeman, K.C. 2000,
BAAS, 197.4115

\end{thebibliography}
\end{document}